\documentclass[twocolumn,showpacs,amsmath,amssymb,pra,superscriptaddress]{revtex4-2}

\usepackage{graphicx}
\usepackage{dcolumn}
\usepackage{bm}
\usepackage{subfigure}
\usepackage{blindtext}

\begin{document}


\title {
Weak non-Landau-type contributions to the diamagnetism of graphite
}

\author {A. V. Nikolaev}

\affiliation{
Skobeltsyn Institute of Nuclear Physics, Moscow
State University, Vorob'evy Gory 1/2, 119234, Moscow, Russia
}

\author {A. V. Bibikov}

\affiliation{
Skobeltsyn Institute of Nuclear Physics, Moscow
State University, Vorob'evy Gory 1/2, 119234, Moscow, Russia
}

\author {M. Ye. Zhuravlev}

\affiliation{St. Petersburg State University, St. Petersburg, 199034, Russia}

\author {L. L. Tao}

\affiliation{School of Physics, Harbin Institute of Technology, Harbin 150001, China}


\begin{abstract}
The exceptionally large diamagnetic susceptibility of a single-crystal graphite in the direction perpendicular to the graphene layers 
is caused by the Landau levels.
However, there are also other contributions to the magnetic susceptibility. In particular, in metals the Pauli paramagnetism is
leading, whereas in dielectrics the diamagnetic Langevin and paramagnetic Van Vleck terms are essential. 
These components of magnetic susceptibility are also present in graphite. 
In this work, we calculate the magnitudes of all these small contributions from first principles.
It is found that the Pauli paramagnetism is negligible (2.3$ \times 10^{-9}$ emu/g)   
whereas the other diamagnetic contribution -- arising from the Langevin and Van Vleck mechanisms -- 
is comparable to the in-plane experimental diamagnetism of graphite and to the diamagnetic response of C$_{60}$ and C$_{70}$ fullerenes.
This diamagnetic contribution proves to be slightly anisotropic ($-2.93 \times 10^{-7}$ emu/g along the $x-$ or $y-$axes
and $-2.24 \times 10^{-7}$ emu/g along the $z-$axis), with the averaged value $-2.7 \times 10^{-7}$ emu/g.
\end{abstract}


\maketitle

\section{Introduction}
\label{sec:int}

Graphite is an important technological material representing a mix of everyday uses, industrial necessity, and strategic value \cite{graphite1,graphite2}.
For example, dry graphite powder is used as a lubricant at extreme temperatures where oil would burn or freeze.
Graphite is a necessary component of lithium-ion batteries because it can reversibly store lithium ions between its layers.
Graphite-derived materials are often used in aerospace, sports equipment, and wind turbine blades.
The most well-known of these materials -- graphene (or few-layer graphene) -- remains a subject of active scientific research,
as it possesses a range of outstanding properties.

In this work we will investigate its magnetic properties which appear in applied external magnetic field $H$ \cite{magprop,Gan,Her,book-NV}.
The task of predicting magnetic properties of solids and nano-structures is of fundamental importance.
In contrast to the problem of obtaining and predicting the electronic structure,
the situation with calculations of magnetic properties remains generally unsolved.
Yet the electron structure of graphite, especially its unusually thin Fermi surface located
very close the angular $KH$-edges of
the Brillouin zone, still attract much attention of researchers \cite{FS-Gru1,FS-Gru2,FS-Schnei}.

The diamagnetic susceptibility of graphite in the direction perpendicular to its planes is the largest among elementary solids \cite{Gan,Her,book-NV}.
The main contribution ($\chi_{Lan}$) to it is due to the Landau mechanism \cite{Lan1,AM}.
Our preliminary results based on the computation of the Fermi surface of graphite and method described in \cite{Nik1} supports this viewpoint.
However, there are other contributions to the magnetic susceptibility $\chi$, namely, the diamagnetic Langevin ($\chi_L$) contribution \cite{VV},
the paramagnetic Van Vleck ($\chi_{VV}$) and Pauli ($\chi_{P}$) contributions \cite{VV,AM}, so that
\begin{eqnarray}
   \chi =  \chi_{Lan} + \chi_{L} + \chi_{VV} + \chi_{P} .
\label{i1}
\end{eqnarray}
The question is: how large are the remaining non-Landau (other than $\chi_{Lan}$) contributions -- $\chi_{L}$, $\chi_{VV}$, and $\chi_{P}$?
Do they constitute a significant fraction of the total $\chi$?
That is the main goal of the present study, which is based on {\it ab initio} band structure calculations of graphite.
To the best of our knowledge in the literature there is no systematic investigation of this problem.

The calculation of the Pauli paramagnetic response $\chi_{P}$ seems straightforward -- it is proportional to the total
density of states (DOS) at the Fermi energy, $N(E_F)$ \cite{AM}. Nevertheless, a serious caveat here is the precise calculation of $N(E_F)$,
which is a difficult task because of the very small number of free carriers in graphite. We describe the problem in more detail in Sec.\ \ref{sub-pauli}.

In contrast to $\chi_{P}$ the calculation of $\chi_{L}$ and $\chi_{VV}$ is a very complicated problem \cite{ML96,MPL,Laskow12,Laskow14,Nik-VV,NikD}. 
In fact, these two contributions should be incorporated in one quantity,
\begin{eqnarray}
   \chi_D =  \chi_{L} + \chi_{VV} ,
\label{i2}
\end{eqnarray}
because the values of individual terms (i.e. $\chi_{L}$, $\chi_{VV}$) depend on the choice of the electromagnetic gauge.
On the other hand, it can be proved (see Ref.\ \cite{NikD}, Appendix~A) that the sum of two terms, $\chi_D$ is
invariant in respect to a gauge transformation.
The first method for the calculation of $\chi_D$ (that is, the magnetic susceptibility of insulators) from first principles was developed by
Mauri and Louie \cite{ML96}.
Since then, several modifications and extensions have been proposed (e.g. \cite{MPL,Laskow12,Laskow14,NikD}), employing various gauge-fixing schemes.
In this work, we follow the method \cite{NikD}
used to calculate $\chi_D$ in diamond from first principles, which involves choosing a single magnetic gauge for the entire crystal.

The paper is organized as follows. In Sec.~\ref{sec:method} we briefly review the method, including explicit expressions 
used to calculate the Langevin diamagnetic contribution in Sec.\ \ref{sub-Lan}, the Van Vleck paramagnetic contribution in Sec.~\ref{sub-VV}, 
and the Pauli paramagnetic response in Sec.~\ref{sub-pauli}. In Sec.~\ref{sub-aver} the procedure of averaging, required to obtain 
the correct symmetry, is discussed in detail, and in Sec.\ \ref{sub:lapw} the technical details of {\it ab initio} calculations are given. 
In Sec.~\ref{sec:RES} the method is applied to graphite, where we calculate $\chi$ for various directions of the applied magnetic field $\vec{H}$. 
The final results are obtained in two steps: in Sec.\ \ref{sub:unit-cell} we discuss calculations of $\chi$, obtained with a single unit cell 
(auxiliary quantities), which are used in the second step in Sec.\ \ref{sub:final} to obtain the final results.
Our conclusions are summarized in Sec.~\ref{sec:con}.

\section{Method}
\label{sec:method}

As mentioned in the introduction, we work with a single choice of magnetic gauge throughout the whole crystal \cite{NikD}.
The gauge origin $\vec{R}_0$ coincides
with the origin of the coordinate system, i.e. $\vec{R}_0=0$, and the focus is on the first ($n=1$) unit cell.
It can be proved that the contribution from another call ($n \neq 1$) is exactly the same as for the $n=1$ unit cell \cite{NikD}.
The vector potential reads as
\begin{eqnarray}
   \vec{A} = \frac{1}{2}\, \vec{H} \times \vec{r} = \frac{1}{2}\, r_{\perp} H\, \vec{e}_{\perp, H} ,
\label{m0}
\end{eqnarray}
where $r_{\perp}$ is the component of the radius vector $\vec{r}$, perpendicular to the direction of the magnetic field $H$, $\vec{n}_H=\vec{H}/H$, 
drawn through the point $\vec{R}_0=0$, 
with the unit vector $\vec{e}_{\perp, H} \sim \vec{r}_{\perp} \times \vec{n}_H$.

\subsection{Langevin diamagnetic response ($\chi_{L}$)}
\label{sub-Lan}

First, we consider the Langevin diamagnetic response $\chi_L$, which has two main contributions -- from the valence and core electrons,
$\chi_{L}$(tot.)$=\chi_L$(val.)$+\chi_L$(core).
The general expression \cite{LanS,LanQ} is
\begin{eqnarray}
   \chi_{L} = -\frac{e^2}{4mc^2} \langle r_{\perp}^2 \rangle ,
\label{m1}
\end{eqnarray}
where $\langle r_{\perp}^2 \rangle$ stands for averaging over the core electron density $\rho_{core}(r)$ or valence electron density $\rho_{val}(r)$
(and $r_{\perp}$ is the radius vector component perpendicular to $H$).
For the averaged value, $\langle r_{\perp}^2 \rangle$, we have,
\begin{eqnarray}
   \langle r_{\perp}^2 \rangle = \int_{V} \rho(r)\, r_{\perp}^2 \; dv ,
\label{m2}
\end{eqnarray}
where the integration is taken over the unit cell region (with the volume $V$).
In the case of core electron shells the electron density $\rho_c$ is spherical and confined by the interior region of MT-spheres
centered at nuclei. In that case the region of integration in Eq.~(\ref{m2}) is simply the MT-spheres.
For the valence electrons ($\rho_{val}$), the integration should be performed over the whole unit cell including the interstitial region.
Thus, the valence contribution consists of two parts, $\chi_L$(val.)$=\chi_L$(IR)$+\chi_L$(MT),
from the interstitial region (IR) and MT-spheres (MT).
Note, that in the interstitial region the value of $r^2_{\perp}$ is directly determined by
the direction of the magnetic field $\vec{n}_H$,
which explicitly changes the integral $\langle r^2_{\perp} \rangle$, Eq.\ (\ref{m2}).

Further, as demonstrated in \cite{NikD}, if MT-sphere is displaced from $O(0,0,0)$ through the translation by the vector $\vec{R}$,
then an additional (off-set) term should be also taken into account, since
\begin{eqnarray}
   \chi^{dia}_{MT,R} = -C_L \langle (\vec{r} + \vec{R})_{\perp}^2 \rangle_c  \nonumber \\
   = \chi^{dia}_{MT,0} - C_L R_{\perp}^2 |Q^{MT}| ,
\label{m3}
\end{eqnarray}
where $R_{\perp}$ is the component of $\vec{R}$ perpendicular to $\vec{H}$,
and $Q^{MT}$ is the total electron charge inside the MT-sphere.
Note, that the second term on the right hand side of Eq.\ (\ref{m3}) (i.e. $- C_L R_{\perp}^2 |Q^{MT}|$) represents
the offset term, appearing because the origin ($O$) does not coincide with the center of MT-sphere.

\subsection{Van Vleck paramagnetic response ($\chi_{VV}$)}
\label{sub-VV}

The Van Vleck paramagnetic contribution $\chi_{VV}$ to the magnetic susceptibility $\chi$ \cite{LanQ}, adapted for the band structure case,
can be written in the following form \cite{NikD}:
\begin{eqnarray}
   \chi_{VV} = 2 \sum_{\vec{k}} w(\vec{k})\, \sum_a^{N_a^o} \, \sum_b^{N_b^{uno}} \; \frac{| \langle b\, | M_H(\vec{k}) | a \rangle |^2}{E_{b}(\vec{k})-E_a(\vec{k})} .
\label{v1}
\end{eqnarray}
Here, the first summation (on $a$) is taken over all occupied bands $N_a^o$ and the second (on $b$) over unoccupied bands $N_b^{uno}$.
In general, in Eq.~(\ref{v1}) one has to integrate over all points $\vec{k}$ belonging to the first Brillouin zone. 
In practice the summation is taken over a set of a representative $k-$points, Sec.\ \ref{sub:lapw}, with corresponding $k-$weights $w(k)$.
For graphite the Fermi energy lies between electron bands with indices $a=8$ and $b=9$, for graphene -- $a=4$ and $b=5$.
Therefore, in Eq.\ (\ref{v1}) $a=1-8$, $N_a^o = 8$ for graphite and $a=1-4$, $N_a^o = 4$ for graphene, and, correspondingly, 
$b \geq 9$ for graphite and $b \geq 5$ for graphene. 
The upper limit for the unoccupied bands, $N_b^{uno}$, is determined by the dimension $N_b$ of the basis set used for the band structure calculation,
$N_b^{uno} = N_b - N_b^o$. For graphite we have $N_b=725$, for graphene $N_b=875$, see Sec.\ \ref{sub:lapw} for more details. 

The electron magnetic operator is given by $M_H \equiv M_{z'}=\mu_B(g_s s_{z'} + L_{z'} )$, where $\mu_B$ is the Bohr magneton and $g_s \approx 2$ is the electron $g-$factor.
Here the $z'$-axis correspond to the direction $\vec{n}_H$ of the external magnetic field $\vec{H}$ introduced earlier.
Note that in the absence of the spin-orbit coupling,
due to the orthogonality of the band states at each $k-$point, we have $\langle b | a \rangle = 0$ ($a \neq b$) and hence $\langle b | s_{z'} | a \rangle = 0$.
Therefore, the contribution to $\chi_{VV}$ stems only from the matrix elements of the orbital momentum $L_H \equiv L_{z'}$, and
\begin{eqnarray}
   \langle b\, | M_H(\vec{k}) | a \rangle = \mu_B \langle b\, | L_H(\vec{k}) | a \rangle ,
\label{v2}
\end{eqnarray}
In practice we obtain the eigenstates $| a \rangle$, $| b \rangle$ from electron band structure calculations.
When calculating the Van Vleck paramagnetic contribution, one must take into account the contribution from the interstitial region (outside the MT spheres) and, 
as a rule, an additional offset term as well, since the magnetic moment depends on the choice of the coordinate origin, which does not necessarily coincide with 
the center of the MT sphere.
Indeed, if the center of the MT-sphere is defined by a vector $\vec{R} \neq 0$, then the offset term appears in the expression for the orbital momentum: 
\begin{eqnarray}
  \langle b | L'_H | a \rangle = \langle b \, | (\vec{r} + \vec{R}) \times \vec{p}\, | a \rangle \cdot \vec{n}_H  \nonumber \\
  = \langle b | L_H | a \rangle + \vec{R}_{\perp} \times \langle b \, | \vec{p}_{\perp}\, | a \rangle \cdot \vec{n}_H ,
\label{v3}
\end{eqnarray}
where $\vec{p}$ is the momentum operator, and $\langle L_H \rangle$ are matrix elements for the orbital momentum in respect to the MT-sphere center.

Matrix elements of full magnetic moment $M_H$ are thus given by
\begin{eqnarray}
   \langle b\, | M_H(\vec{k}) | a \rangle &=& \mu_B (\langle b\, | L_H^{MT}(\vec{k}) | a \rangle +
   \langle b\, | L_H^{MT,\, off}(\vec{k}) | a \rangle  \nonumber \\
   & &+ \langle b\, | L_H^{IR}(\vec{k}) | a \rangle).
\label{v4}
\end{eqnarray}
Here $L_H^{MT}$ is the orbital moment from the MT-spheres (with the magnetic moment counted from the sphere center), $L_H^{IR}$
is the contribution from the interstitial region, while $L_H^{MT,\, off}$ is the offset term.

Finally, we recall that only the sum of the Langevin and Van Vleck terms $\chi_D$, Eq.\ (\ref{i2}), is the gauge invariant \cite{NikD}.

\subsection{Averaging procedure for the resulting susceptibilities ($\chi_L$, $\chi_{VV}$, $\chi_D$)}
\label{sub-aver}

\begin{figure}
\resizebox{0.35\textwidth}{!} {
\includegraphics{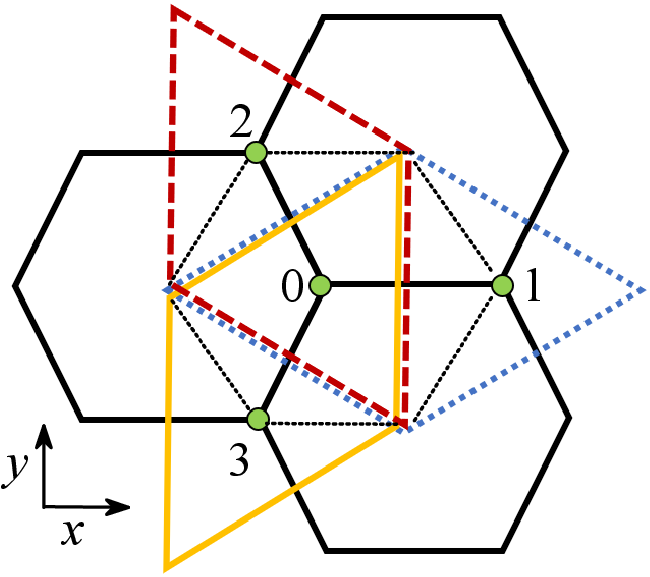}
}

\caption{
Three distinct unit cells of graphite layer (graphene), used for averaging their contributions to obtain
the resulting magnetic susceptibility $\chi$, adapted to the layer symmetry:
(1) with atoms 0 and 1 (blue dot line),
(2) with atoms 0 and 2 (red dashed line), (3) with atoms 0 and 3 (yellow solid line).
The hexagon centered at atom $0$, shown by a dotted black line, represents a unit cell possessing full symmetry.
} \label{fig1}
\end{figure}

One of the most important consequences of the induced magnetization, caused by the external magnetic field $H$, is
the emergence of a new crystalline symmetry, which takes into account the direction $\vec{n}_H$ of $\vec{H}$. 
For example, the application of $H$ in the direction $\vec{n}_H$ collinear to the [100] axis 
changes the symmetry of graphite layer from the three-fold rotation symmetry ($C_3$) about the $z-$axis at the carbon atom $0$ in Fig.\ \ref{fig1} to 
the two-fold one ($C_2$) about the $x-$axis.  
As a result, for a general direction of $H$ the calculation of $\chi_{VV}$ in Eq.\ (\ref{v1}) requires the integration over the whole Brillouin zone,
the reduction to the irreducible part ($1/24$) of the Brillouin zone is not possible.

%
%
The other important consequence is that the unit cell in direct lattice should possess the full crystal symmetry \cite{NikD}. 
For the graphite layer, this would require a specific choice of a hexagonal unit cell, shown by the thin dotted black line in Fig. \ref{fig1}.
(Atom $0$ is located at the center, and atoms $1, 2,$ and $3$ are at the vertices.) 
However, in practice such a unit cell is inconvenient for band structure calculations.  
The alternative solution to this problem is to use a few unit cells having different orientation in respect to $H$, 
and then average their results.
Such set of three unit cells (each with two carbon atoms) for a graphite layer is shown in Fig.\ \ref{fig1}.   
In general, calculated values $\chi_i$ of three unit cells ($i=1,2,3$) differ from each other and do not have the full symmetry.  
However, the correct full symmetry is restored after averaging over them,  
\begin{eqnarray}
   \chi(\vec{n}_H) = \frac{1}{3} (\chi_1(\vec{n}_H) + \chi_2( \vec{n}_H) + \chi_3(\vec{n}_H)) .
\label{a1}
\end{eqnarray}
Here $\chi_i(\vec{n}_H)$ ($i=1,2,3$) is the magnetic susceptibility ($\chi_L$, $\chi_{VV}$ or $\chi_D$) obtained with the first, second or third unit cell, 
shown in Fig.\ \ref{fig1}.
Note, that $\chi_2(\vec{n}_H) = \chi_1(\hat{R}\, \vec{n}_H)$, where $\hat{R}$ is the operation of rotation by the angle
$2\pi/3$ about the $z-$axis. Analogously, $\chi_3(\vec{n}_H) = \chi_1(\hat{R}^{-1}\, \vec{n}_H)$
where $\hat{R}^{-1}$ is the operation of rotation by the angle
$-2\pi/3$ about the $z-$axis.

Therefore, the task of obtaining the correct values of $\chi_L$, $\chi_{VV}$ and $\chi_D=\chi_L+\chi_{VV}$ 
is done in two steps. First, we calculate $\chi_L$, $\chi_{VV}$ and $\chi_D$ for a single unit cell (number one in Fig.\ \ref{fig1}).  
This step is described in detail in Sec.\ \ref{sub:unit-cell} below. Second, in Sec.\ \ref{sub:final} we effectively average the obtained 
values over three distinct unit cells, Eq.\ (\ref{a1}) and obtain the final results.

\subsection{Pauli paramagnetic response}
\label{sub-pauli}

The Pauli paramagnetic susceptibility is given by
\begin{eqnarray}
   \chi_P = \mu_B^2 \, N(E_F) ,
\label{p1}
\end{eqnarray}
where $\mu_B$ is the Born magneton and $N(E_F)$ is the density of states of graphite at the Fermi energy.
The problem is to calculate $N(E_F)$ with a high precision, which is a difficult task because
the number of electrons and holes in graphite is vanishingly small ($\sim 10^{-5}$ per unit cell).
The characteristic radius of the Fermi surface at the $KH$-edge of the Brillouin zone (BZ)
is only 0.0005-0.005 (in a.u.$^{-1}$), and to access this size a huge number (more than $10^6$) of $k$-points
in BZ is required.
Under such conditions the regular methods used in computational solid state physics give $N(E_F)$ with a large error.
Even the use of the tetrahedron method \cite{tet} with its linear interpolation between the $k$-points
cannot save the situation.

For that reason, for highly accurate estimations (last line in Table \ref{tab6} below) we have adopted a special approach.
First, we have determined the Fermi surface $S_{FS}(E \equiv E_F)$ of graphite, and then
calculated the DOS at $E_F$ by means of the standard integral relation \cite{AM}
\begin{eqnarray}
    N(E_F) = \int_{k \in S_{FS}} \frac{dS}{4\pi^3} \frac{1}{|\nabla E(\vec{k})|} ,
 \label{p3}
\end{eqnarray}
where the integration is performed over the Fermi surface $S_{FS}$ in BZ and $\nabla E(\vec{k})$
is the gradient of the dispersion law $E(\vec{k})$, computed at the Fermi surface.

\subsection{Computation details}
\label{sub:lapw}

In our work we have used the full linear augmented plane wave (FLAPW) method \cite{blapw,wien2k} (implemented in Ref.\ \cite{Nik2}),
which is one of the most precise all-electron method for band structure calculations.
$\chi_{L}$, $\chi_{VV}$ and $\chi_{D}$ were computed with the core and valence electron densities
obtained as a result of the FLAPW self-consistent procedure.
We used the basis vectors of the direct and reciprocal lattices as defined in Ref.\ \cite{BC}.

Calculations of Pauli paramagnetism $\chi_P$, DOS at the Fermi energy $N(E_F)$ and the Fermi surface have been performed
for the bulk (3D) graphite material.
Calculation of the Langevin ($\chi_L$) and Van Vleck ($\chi_{VV}$) contributions have been done for a quasi-2D structure, or graphene. 
For such calculations we have used the AA-graphite supercell with a very large lattice constant along the $z$-axis,
perpendicular to the graphite layer.
The main reason for that is the instability of $\chi_{VV}$ related to Eq.\ (\ref{v1}).
The value of the sum in Eq.\ (\ref{v1}) is very sensitive towards the $k$-point grid used for calculations,
especially in the region close to the $KH$-edge of BZ (or close to the Fermi surface), because there the occupied $E_a(\vec{k})$ bands
lie very close to the unoccupied bands $E_b(\vec{k})$, and $E_a(\vec{k}) \approx E_b(\vec{k})$.
For this region the denominator in Eq.\ (\ref{v1}) becomes very small, and, as a result, each $k$-point leads to a very large contribution.
As argued in Sec.\ \ref{sub-pauli}, it is impossible to reach convergence of the sum in Eq.\ (\ref{v1}) using a reasonable number of $k$-points.
By reducing the problem to a quasi-two-dimensional structure with the sharp division between the occupied and unoccupied bands we have stabilized 
the summation process in Eq.\ (\ref{v1}) and ensured the convergence of $\chi_{VV}$. 

The technical parameters of numerical calculations are the following.
For the 3D case of graphite the number of augmented plane waves was 725 with the wave vectors $\vec{K}_j$ satisfying the condition $R_{MT} K_j \leq 7.6$.
The number of $k-$points in the irreducible part ($1/24$) of the first BZ was 2352 ($\sim$56000 in the whole BZ), 
the MT-sphere radius was $R_{MT} = 1.342$~a.u.
We used the tetrahedron method for the linear interpolation of
energy between $k-$points \cite{tet}.
For calculation of the exchange-correlation potential and the exchange-correlation energy contribution within the DFT approach,
we have used two different variants: (1) the Perdew-Burke-Ernzerhof (PBE) scheme \cite{PBE} of the generalized-gradient approximation (GGA),
and (2) the local density approximation (LDA) with the standard ($V_{exc} \sim - \rho^{1/3}$) exchange \cite{Dir} and the PW-correlation \cite{PWcorr}.
The number of radial points inside the MT region was 755.
In this work we adopted experimental lattice constants of graphite: $a_{latt}=2.461$~{\AA}, $c_{latt}=6.708$~{\AA}.

For the quasi-2D case the number of augmented plane waves was 875 with the wave vectors $\vec{K}_j$ satisfying the condition $R_{MT} K_j \leq 7.41$.
The number of $k-$points in the irreducible part ($1/24$) of the first BZ was 512 ($\sim$12200 in the whole BZ), the MT-sphere radius was $R_{MT} = 1.342$~a.u.
In that case the primitive unit cell of the AA-graphite (only two carbon atoms) with an artificially enlarged lattice constant $c_{latt}=9$~{\AA} was adopted.
Test calculations of the in-plane equilibrium lattice constants $a_{latt}$ gave $a_{latt}^{LDA}=2.476$~{\AA}, $a_{latt}^{GGA}=2.499$~{\AA},
in good correspondence with experimental data (2.461~{\AA}).
We recall that for the Van Vleck calculations all 875 eigenvectors (including occupied and unoccupied states) at each $k$-point of BZ are required, Eq.\ (\ref{v1}).

Finally, it is worth noting that the application of a magnetic field alters the crystal symmetry — manifesting as an increase in the irreducible part of BZ
— and that, in the general case, the summation in Eq.\ (\ref{v1}) for $\chi_{VV}$ requires calculations performed over the entire BZ.
Unlike $\chi_{VV}$, the Langevin diamagnetic contribution $\chi_L$ is determined by the electron density distributions of valence ($\rho_{val}$)
and core ($\rho_{c}$) electrons, Eqs.\ (\ref{m1})–(\ref{m3}), which can be obtained during a standard band structure calculation
using only the irreducible (1/24) part of BZ.

\section{RESULTS}
\label{sec:RES}

As discussed in Sec.\ \ref{sub-aver}, the task of obtaining the magnetic susceptibility of $\chi_D=\chi_L+\chi_{VV}$
includes two steps. First, we calculate the Langevin term $\chi_L$, the Van Vleck term $\chi_{VV}$ and their sum $\chi_D$ with a single primitive unit cell -- 
first cell (dotted blue line) in Fig.\ \ref{fig1}, with the integration in Eq.\ (\ref{v1}) performed over the whole Brillouin zone. 
This part of calculations is described in Sec.\ \ref{sub:unit-cell}. 
We emphasize that the results presented in Sec.\ \ref{sub:unit-cell} are only auxiliary since they reflect the symmetry of the chosen unit cell. 

The second step, described in detail in Sec.\ \ref{sub:final} below, takes into account the averaging procedure, Eq.\ (\ref{a1}), 
over three unit cells, shown in Fig.\ \ref{fig1}, Sec.\ \ref{sub-aver}, and leads to final correct results, 
adapted to the crystal symmetry.

In Sec.\ \ref{sub:weak-Pauli} we discuss the weak Pauli paramagnetism of graphite.

\subsection{Auxiliary calculations of $\chi_{VV}$, $\chi_{L}$ and $\chi_{D}$ -- \\ for a single unit cell}
\label{sub:unit-cell}

In this subsection we present results obtained with the first unit cell in Fig.\ \ref{fig1}, which do not possess 
the correct crystal symmetry. The full symmetry will be restored in Sec.\ \ref{sub:final} by means of the averaging, Eq.~(\ref{a1}). 

First, we have calculated the Langevin and Van Vleck contributions for three main directions of magnetic field $H$: [1,0,0], [0,1,0] 
and [0,0,1], Table \ref{tab1}. 
%
\begin{table}
\caption{
$\chi_{L}$, $\chi_{VV}$, and their sum $\chi_{D}$ for selected directions $\vec{n}_H$ of $\vec{H}$, 
calculated with the first unit cell in Fig.\ \ref{fig1}. Last line (aver) stands for the averaged values.
All $\chi$ are mass values, in units $10^{-7}$~emu/g.
\label{tab1} }
\begin{ruledtabular}
\begin{tabular}{l   r  r  r   r  r  r}
       &   \multicolumn{3}{c}{$\chi$, LDA} & \multicolumn{3}{c}{$\chi$, GGA}   \\
 $\vec{H} \parallel$ & $\chi_{L}$  & $\chi_{VV}$  & $\chi_{D}$  & $\chi_{L}$  & $\chi_{VV}$  & $\chi_{D}$  \\
\tableline
  $[1,0,0]$    &  -9.225 &  7.427 & -1.799 &  -9.202 &  7.406 & -1.796 \\
  $[0,1,0]$    & -11.159 &  7.102 & -4.057 & -11.152 &  7.080 & -4.072 \\
  $[0,0,1]$    & -12.441 & 10.184 & -2.257 & -12.443 & 10.202 & -2.241 \\
  aver         & -10.942 &  8.238 &  -2.704 & -10.932 &  8.229 & -2.703 \\
\end{tabular}
\end{ruledtabular}
\end{table}
%
\begin{table}
\caption{
Contributions to $\chi_L$ along the $x$, $y$ and $z$ directions of $H$ (GGA-PBE calculations for the first unit cell in Fig.~\ref{fig1}):
from the valence (val) and core electrons, from the region inside MT-spheres (MT) and from the interstitial region (IR). 
All values in units $10^{-7}$~emu/g.
\label{tab2} }
\begin{ruledtabular}
\begin{tabular}{c  r  r  r  }
  contributions   &  $[1,0,0]$ &  $[0,1,0]$ &  $[0,0,1]$   \\
\tableline
  $\chi_L$        &  -9.202  & -11.152  & -12.443  \\
  val (all)       &  -9.071  & -11.021  & -12.312  \\
  val (IR)        &  -7.702  &  -9.652  & -10.804 \\
  val (MT)        &  -1.369  &  -1.369  &  -1.508  \\
  core (MT)       &  -0.131  &  -0.131  &  -0.131  \\
  val $+$ core (MT) &  -1.499  &  -1.499  &  -1.639  \\
  off-site        &    0     &  -3.797  &  -3.797  \\
\end{tabular}
\end{ruledtabular}
\end{table}
Various partial contributions to $\chi_{L}$, discussed in Sec.\ \ref{sub-Lan}, are given in Table \ref{tab2}. 
Note, that the largest partial contribution ($\chi_{L,IR}$) is due to the interstitial region (IR). 
The order of $\chi_{L,IR}$ for three directions of $H$, i.e. $\chi_{L,IR}[1,0,0] < \chi_{L,IR}[0,1,0] < \chi_{L,IR}[0,0,1]$, 
reflects the shape of the first unit cell, Fig.\ \ref{fig1}. Although the distribution the electron density $\rho(r)$ 
within the cell remains the same, $r_{\perp}$ depends on the choice of $\vec{n}_H$. 
Only in spherically symmetric case of core electrons there is no dependence on $\vec{n}_H$. 
The off-site term, Eq.\ (\ref{m3}), discussed in Sec.\ \ref{sub-Lan}, is zero for [1,0,0] 
since for this direction $R_{\perp} = 0$. 

The most important conclusion is that $\chi_{VV}$, $\chi_{L}$ and $\chi_{D}$ are anisotropic in respect to the direction $\vec{n}_H$ of magnetic field. 
Therefore, each of them can be considered as a function $\chi(\vec{n}_H)$ of  polar angles $\vec{n}_H \equiv (\theta, \varphi) \equiv \Omega$, 
and can be expanded in multipolar series in terms of real spherical harmonics \cite{BC},
\begin{eqnarray}
   \chi(\theta, \varphi) = \chi_0 + \sum_{L \ne 0,\tau} \chi_{L,\tau} \, Y_{L,\tau}(\theta, \varphi), \quad
  \label{r1}
\end{eqnarray}
where $\chi_0$ is a constant, giving the averaged over $\vec{n}_H$ value of magnetic susceptibility,
whereas $Y_{L,\tau}(\theta, \varphi)$ are real spherical harmonics \cite{BC}. Here $L$ is the orbital index (in the sum $L \neq 0$)
and $\tau=(m,c)$ or $(m,s)$, i.e. $cos$ or $sin$-type of the $\varphi$-dependence.
The explicit form of $Y_{L,\tau}(\theta, \varphi)$ can be found e.g. in \cite{BC}),
The coefficients of expansion in Eq.\ (\ref{r1}) $\chi_{L,\tau}$ are given by
\begin{eqnarray}
  \chi_{L,\tau} = \int_{\Omega} d \Omega \, \chi(\theta, \varphi) \, Y_{L,\tau}(\theta, \varphi)  \nonumber \\
     \rightarrow \sum_i w_i \chi(\theta_i, \varphi_i) \, Y_{L,\tau}(\theta_i, \varphi_i) ,
  \label{r3}
\end{eqnarray}
where the integration on $\Omega=(\theta, \varphi)$ is replaced by summation on points $(\theta_i, \varphi_i)$ with weights $w_i$.
To perform the summation  in Eq.\ (\ref{r3}), we have used the 50 point Lebedev surface grid \cite{Leb1,Leb2}, which is enough to extract
the coefficients up to $L=4$. (The Lebedev grid includes points $\{\pm 1,0,0\}$, $\{0,\pm 1, \pm1\}$, $\{\pm 1,\pm 1,\pm 1\}$, $\{\pm a, \pm a, \pm b\}$,
where $a=0.30151$, $b=0.90453$, and the others obtained from them by cyclic permutations.)
As a result, we find that only the real spherical harmonics $Y_{2,0}$, $Y_{2,2c}$, and $Y_{2,2s}$ yield non-zero coefficients $\chi_{2,0}$, $\chi_{2,2c}$, and $\chi_{2,2s}$, whereas the other possible coefficients (with $L=1, 3, 4$) are practically zero.
Therefore, the general expansion of $\chi$, Eq.\ (\ref{r1}), reduces to
\begin{eqnarray}
   \chi^A(\theta, \varphi) & = & \chi^A_0 + \chi^A_{2,0} Y_{2,0}(\theta, \varphi) + \chi^A_{2,2c} Y_{2,2c}(\theta, \varphi)  \nonumber \\
                  &   & + \chi^A_{2,2s} Y_{2,2s}(\theta, \varphi) . 
  \label{r2}
\end{eqnarray}
The calculated values of the coefficients 
$\chi^A_{2,0}$, $\chi^A_{2,2c}$ and $\chi^A_{2,2s}$ are quoted in Table~\ref{tab1}.
The relations (\ref{r2}) for $\chi^A_{L}$, $\chi^A_{VV}$, and $\chi^A_D$, with the coefficients $\chi^A_{L,\tau}$ listed in Table~\ref{tab3}, 
provide a complete description of their angular dependencies.
%
\begin{table}
\caption{
Auxiliary coefficients  $\chi^A_{L,\tau}$ of expansion in real spherical harmonics, Eq.~(\ref{r2}), 
$\tau = m,c$ or $m,s$ \cite{BC},
for $\chi_{L}$, $\chi_{VV}$, and $\chi_{D}$, calculated with the 50 point Lebedev grid with the first unit cell in Fig.\ \ref{fig1}.
$\chi^A(0)$ is also the averaged value.
All values in units $10^{-7}$~emu/g.
\label{tab3} }
\begin{ruledtabular}
\begin{tabular}{l  r  r  r   r  r  r}
     &   \multicolumn{3}{c}{LDA} & \multicolumn{3}{c}{GGA}   \\
      & L  & VV  & D   & L  & VV  & D   \\
\tableline
  $\chi^A(0)$ (aver)  & -10.942 &  8.238 &  -2.704 & -10.932 &  8.229 & -2.703 \\
  $\chi^A(2,\, 0)$    &  -2.377 &  3.085 &   0.708 &  -2.395 &  3.128 &  0.733 \\
  $\chi^A(2,\, 2c)$    &   1.770 &  0.297 &   2.066 &   1.785 &  0.299 &  2.084 \\
  $\chi^A(2,\, 2s)$    &   3.085 & -2.917 &   0.168 &   3.091 & -2.932 &  0.158 \\

\end{tabular}
\end{ruledtabular}
\end{table}

\subsection{Symmetry adapted final results for\\ $\chi_{VV}$, $\chi_{L}$ and $\chi_{D}$}
\label{sub:final}

%
\begin{table}
\caption{
Coefficients of expansion $\chi_0$, $\chi_{2,0}$ in real spherical harmonics, Eq.~(\ref{r6}), 
for diamagnetic (Langevin) contribution $\chi_{L}$, for paramagnetic (Van Vleck) contribution $\chi_{VV}$, and their
sum $\chi^{D}$, calculated with the 50 point Lebedev grid.
$\chi_0$ is also the averaged value.
All values in units $10^{-7}$~emu/g.
\label{tab4} }
\begin{ruledtabular}
\begin{tabular}{l  r  r  r   r  r  r}
     &   \multicolumn{3}{c}{LDA} & \multicolumn{3}{c}{GGA}   \\
       & dia  & para  & sum  & dia  & para  & sum  \\
\tableline
  $\chi(0)$ (aver)  & -10.942 &  8.238 &  -2.704 & -10.932 &  8.229 & -2.703 \\
  $\chi(2,\, 0)$    &  -2.377 &  3.085 &   0.708 &  -2.395 &  3.128 &  0.733 \\
\end{tabular}
\end{ruledtabular}
\end{table}

In this subsection on the basis of auxiliary calculations, described earlier in Sec.\ \ref{sub:unit-cell}, we obtain 
final results for $\chi_{L}$, $\chi_{VV}$, and their sum $\chi_{D}$. 

\begin{figure}
\resizebox{0.35\textwidth}{!} {
\includegraphics{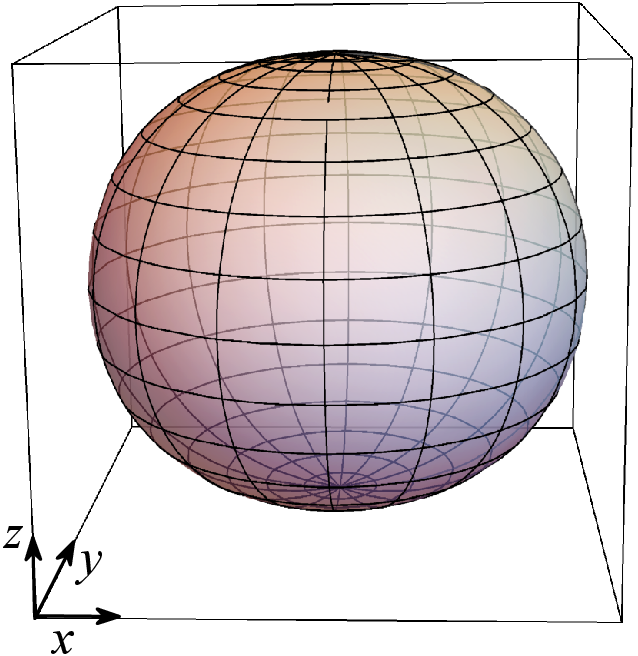}
}

\caption{
Anisotropic (slightly oblate along the $z$-axis) ellipsoid of the diamagnetic susceptibility $|\chi_D|$ ($\chi_D=\chi_L+\chi_{VV} < 0$)
as a function of the direction ($\vec{n}_H$) of the applied magnetic field $H$, 
GGA-PBE calculations.
} \label{fig2}
\end{figure}

The auxiliary quantities $\chi_{L,1}$, $\chi_{VV,1}$, and $\chi_{D,1}$ -- as indicated by the subscript $1$ -- 
were calculated for the first unit cell (Fig.\ \ref{fig1}).
The final results are obtained by the averaging, Eq.\ (\ref{a1}), over three distinct unit cells, shown in Fig.\ \ref{fig1}.  
Knowing the angular dependence of $\chi^A_1(\theta, \varphi)$, Eq.\ (\r2), (where $\chi^A_1=\chi^A_{L,1}$, $\chi^A_{VV,1}$, $\chi^A_{D,1}$) 
the final susceptibility is given by   
\begin{eqnarray}
   \chi(\theta, \varphi) & = & \frac{1}{3}(\chi^A_1(\theta, \varphi) + \chi^A_1(\theta, \varphi + 2\pi/3) \nonumber \\ 
                            & & + \chi^A_1(\theta, \varphi - 2\pi/3)) .
\label{r4}
\end{eqnarray}
The averaging introduces new expansion coefficients $\chi_0$, $\chi_{2,0}$, $\chi_{2,2c}$, $\chi_{2,2s}$ in place of 
the auxiliary coefficients $\chi^A_0$, $\chi^A_{2,0}$, $\chi^A_{2,2c}$, $\chi^A_{2,2s}$, obtained for the first unit cell (Sec.~\ref{sub:unit-cell}).
We arrive at 
\begin{subequations}
\begin{eqnarray}
   & & \chi_0 = \chi^A_0 , \quad  \chi_{2,0} = \chi^A_{2,0} ,  \label{r5a}  \\
   & & \chi_{2,2c} = \chi_{2,2s} = 0 .  \label{r5b}
\end{eqnarray}
\end{subequations}
Thus, the resulting angular dependence of $\chi_{L}$, $\chi_{VV}$ and their sum $\chi_{D}$ is rather simple,
\begin{eqnarray}
   \chi(\vec{n}_H) = \chi_0 + \chi_{2,0}\, Y_{2,0}(\vec{n}_H) , 
  \label{r6}
\end{eqnarray}
where the corresponding coefficients of expansion are listed in Table \ref{tab4}. 
The dependence of $\chi_D(\vec{n}_H)$ on the direction of the magnetic field $\vec{n}_{H}$ is visualized in Fig.~\ref{fig2}.
%
%
$\chi_{L}$, $\chi_{VV}$ and $\chi_{D}$, being independent of $\varphi$, possess the rotational symmetry around the $z-$axis.
In Table \ref{tab5} we quote magnetic susceptibilities for $x-$, $y-$ and $z-$axis. 
%
\begin{table}
\caption{
Diamagnetic Langevin contribution $\chi_{L}$, paramagnetic Van Vleck contribution $\chi_{VV}$, and their
sum $\chi_{D}$ for $\vec{H}$ aligned along $x-$, $y-$ and $z-$axis. Last line (aver) gives averaged values.
All $\chi$ are mass values, in units $10^{-7}$~emu/g.
\label{tab5} }
\begin{ruledtabular}
\begin{tabular}{l   r  r  r   r  r  r}
       &   \multicolumn{3}{c}{$\chi$, LDA} & \multicolumn{3}{c}{$\chi$, GGA}   \\
 $\vec{H} \parallel$ & $\chi_{L}$  & $\chi_{VV}$  & $\chi_{D}$  & $\chi_{L}$  & $\chi_{VV}$  & $\chi_{D}$  \\
\tableline
  $x$ or $y$   & -10.192 &  7.265 & -2.928 & -10.177 &  7.243 & -2.934 \\
  $z$          & -12.441 & 10.184 & -2.257 & -12.443 & 10.202 & -2.241 \\
  aver         & -10.942 &  8.238 & -2.704 & -10.932 &  8.229 & -2.703 \\
\end{tabular}
\end{ruledtabular}
\end{table}

\subsection{Weak Pauli paramagnetism}
\label{sub:weak-Pauli}

In contrast to the Langevin and Van Vleck contributions, the Pauli paramagnetism $\chi_P$, Eq.\ (\ref{p1}), 
is independent of the direction of magnetic field. 
As discussed in Sec.\ \ref{sub-pauli}, the main problem is to determine the density of states at the Fermi energy with high precision,
which is a difficult task because of vanishingly small amount of electrons and holes in graphite ($\sim 10^{-5}$ per unit cell). 
We have done it by two different ways. 
First, we use a standard band structure method with the linear interpolation of energies between neighboring points -- 
representing the tetrahedron method \cite{tet}. 
However, since the Fermi surface of graphite is very close to the $KH-$edge of the Brillouin zone, this approach
requires a very large number of $k-$points. The largest number we used is 10164 $k$-points in the (1/24) part of BZ, 
and the resulting densities of states for LDA and GGA is listed in Table \ref{tab6} as $N^S(E_F)$. 
%
\begin{table}
\caption{
Calculated density of states $N(E_F)$ and the paramagnetic Pauli contribution $\chi_P$.
$N^S(E_F)$ is the standard band structure calculation with 10164 $k$-points in the (1/24) part of BZ, and
$N^I(E_F)$ is precise value obtained by integration over the Fermi surface, Eq.\ (\ref{p1}).
$\chi_P$, computed with $N^I(E_F)$, is in units $10^{-7}$~emu/g, $N(E_F)$ -- in (states/eV) per unit cell.
\label{tab6} }
\begin{ruledtabular}
\begin{tabular}{c  r  r  r }
               & LDA   & \multicolumn{2}{c}{GGA} \\
               &       &    PBE  &   PBEsol \cite{PBEsol} \\
\tableline
  $N^S(E_F)$   & 0.00533  &  0.00445 &  0.00500  \\
  $N^I(E_F)$   & 0.00353  &  0.00337 &  0.00348  \\
  $\chi_P$     & 0.02377  &  0.02269 &  0.02340  \\
\end{tabular}
\end{ruledtabular}
\end{table}

In the second approach we used Eq.\ (\ref{p3}), which is very precise, but requires the explicit use of the Fermi surface. 
The construction of the Fermi surface is a difficult task, which will be described in detail elsewhere.
Here we reproduce the calculated with Eq.\ (\ref{p3}) value of density of states in Table \ref{tab6} as $N^I(E_F)$. 
We consider $N^I(E_F)$ as the benchmark result, since the error in integration in Eq.\ (\ref{p3}) is very small. 
The resulting values of $\chi_P$ in Table \ref{tab6} are obtained with $N^I(E_F)$.  
The comparison between $N^I(E_F)$ and $N^S(E_F)$ demonstrate that the standard band structure calculation still overestimates 
$N(E_F)$ by 30\%. 
Our experience shows that obtaining a more accurate value requires even more $k$-points in the standard scheme. 

An analysis of the data in Table \ref{tab6} shows that the calculated values of $\chi_P$ are very small -- 
two orders of magnitude lower than the values of $\chi_{VV}$, $\chi_{L}$, and $\chi_{D}$ discussed in Sec.\ \ref{sub:final} -- 
and can safely be neglected in comparison with the latter.

\section{Conclusions}
\label{sec:con}

We have examined minor contributions to the magnetic susceptibility of graphite $\chi$ -- those arising from the Langevin diamagnetism $\chi_L$
and the Van Vleck paramagnetism $\chi_{VV}$. 
All these contributions -- as well as the very weak Pauli paramagnetism $\chi_P$ -- have been calculated from the first principles 
using LDA and GGA (PBE) variants of DFT in Moscow-LAPW code \cite{Nik2}. 
Both variant yield very close results.

The sum of $\chi_L$ and $\chi_{VV}$ results in diamagnetic response $\chi_D = \chi_L + \chi_{VV} < 0$, which depends 
on the direction $\vec{n}_H$ of the applied magnetic field $\vec{H}$. 
The expansion of $\chi_L$, $\chi_{VV}$, and $\chi_D$ in terms of real spherical harmonics leads to equation (\ref{r6}), 
the coefficients of which, $\chi_0$ and $\chi_{2,0}$, are listed in Table \ref{tab4}.
$\chi_L$, $\chi_{VV}$ and $\chi_D$ as functions of $\vec{n}_H$ have rotational symmetry around the $z-$axis,
$\chi_D(\vec{n}_H)$ is plotted in Fig.\ \ref{fig2}. 
Values of $\chi_L$, $\chi_{VV}$ and $\chi_D$ when $\vec{H}$ is aligned along $x-$, $y-$ or $z-$axes are 
reproduced in Table~\ref{tab5}. 
These data demonstrate a weak anisotropy ($\sim 17$\%) of $\chi_D$ with respect to $\vec{n}_H$, 
where $|\chi_D[0,0,1]| < |\chi_D[1,0,0]| = |\chi_D[0,1,0]|$. 

The averaged value for $\chi_D$ ($-2.7 \times 10^{-7}$~emu/g) is approximately half of the experimental in-plane (fields parallel to the layers) 
magnetic susceptibility of graphite ($-5$/$-5.5 \times 10^{-7}$~emu/g \cite{Gan,Her}) 
and comparable to the experimental diamagnetic response for fullerenes C$_{60}$ ($-3.1 \times 10^{-7}$~emu/g \cite{Her}, $-3.5 \times 10^{-7}$~emu/g \cite{Had,Ruo}) 
and C$_{70}$ ($-5.9 \times 10^{-7}$~emu/g \cite{Had}).

In contrast to $\chi_L$, $\chi_{VV}$ and $\chi_D$, the Pauli paramagnetism $\chi_P$ is invariant of the direction of $\vec{H}$. 
The calculated value ($2.3 \times 10^{-9}$~emu/g) is very small (less than 1\%) even in comparison with $\chi_D$, Table \ref{tab6}.  
This is due to the very low density of states at the Fermi energy of graphite ($3.4 - 3.5 \times 10^{-2}$ states/eV per unit cell).

\acknowledgements

The study was carried out within the framework of a state assignment for Lomonosov Moscow State University (A.V. Nikolaev and A.V. Bibikov, Topic 8.4)
and St. Petersburg State University (M.Ye. Zhuravlev). 



\end{document}